\documentstyle[11pt,newpasp,twoside,epsfig]{article}
\def\edcomment#1{\iffalse\marginpar{\raggedright\sl#1\/}\else\relax\fi}
\reversemarginpar

\begin{document}

\title{SAURON dynamical modeling of NGC~2974}

\author{Davor Krajnovi\'c$^{1}$, Michele Cappellari$^{1}$, Eric
Emsellem$^{2}$,\\ Richard McDermid$^{1}$, P. Tim de Zeeuw$^{1}$}
\affil{$^{1}$ Sterrewacht Leiden, P.O.Box 9513, 2300 RA Leiden, The
Netherlands}
\affil{$^{2}$ CRAL, 9 Av. Charles-Andre, 69230 Saint Genis-Laval, France}

\section{Introduction}
NGC~2974 is a nearby field elliptical galaxy classified as E4 with
detected HI and H$\alpha$ gas apparently distributed in a disk (Kim et
al.\ 1988; Buson et al.\ 1993). An earlier study by Cinzano \& van der
Marel (1994) suggested the galaxy harbors an embedded stellar disk not
visible in the photometry. The galaxy was observed with the
integral-field spectrograph {\tt SAURON} (Bacon et al. 2001) mounted
on the 4.2m WHT on La Palma, as a part of the survey of a
representative sample of nearby E, S0 and Sa galaxies (de Zeeuw et
al. 2002). With this individual study we explore the dynamical
structure of the object, give a quantitative analysis of the embedded
disk structure and investigate the distribution and dynamics of the
gas in NGC~2974. A similar analysis is being performed on the galaxies
of the sample, providing constraints on galaxy formation scenarios.
\looseness=-2

\vspace*{-0.1cm}
\section{Stellar dynamical model}
The {\tt SAURON} observations of emission-lines are consistent with
the gas being distributed in a circular disk, except in the inner
3\arcsec\ where a bar-like perturbation was detected (Emsellem,
Ferruit, \& Goudfrooij, 2003). By assuming the gas is in a thin disk
and fitting to the two-dimensional kinematics we are able to constrain
the inclination of the disk to $i \sim 60^{\circ}$. Adopting that
inclination we construct an axisymmetric model for the galaxy stellar
body using a three-integral orbit superposition Schwarzschild modeling
technique adapted for integral field data (e.g. Cretton et al.\ 1999,
Cappellari et al. 2002, Verolme et al 2002). \looseness=-2

The best fitting model (Fig.~1) is able to reproduce all the features
of the observed stellar kinematics.  Analyzing the orbital
distribution of the model we find two clearly distinct kinematic
components. One component consists of high angular momentum orbits,
producing a flattened density distribution, and the other component is
rounder and rotates slowly. We associate the former component with a
stellar disk and the latter with the spheroid. It is possible to
perform a disk-bulge decomposition in phase space and estimate their
mass fraction which turns out to be about 1:3.

We construct a dynamical model for the gas with the same potential and
inclination used for the stellar model. Solving the Jeans equations in
the limit of the asymmetric drift approximation we are able to
reproduce the large scale gas kinematics providing an independent
consistency test of the potential used for the stellar modeling.

\begin{figure}[!t] 
 \epsfig{figure=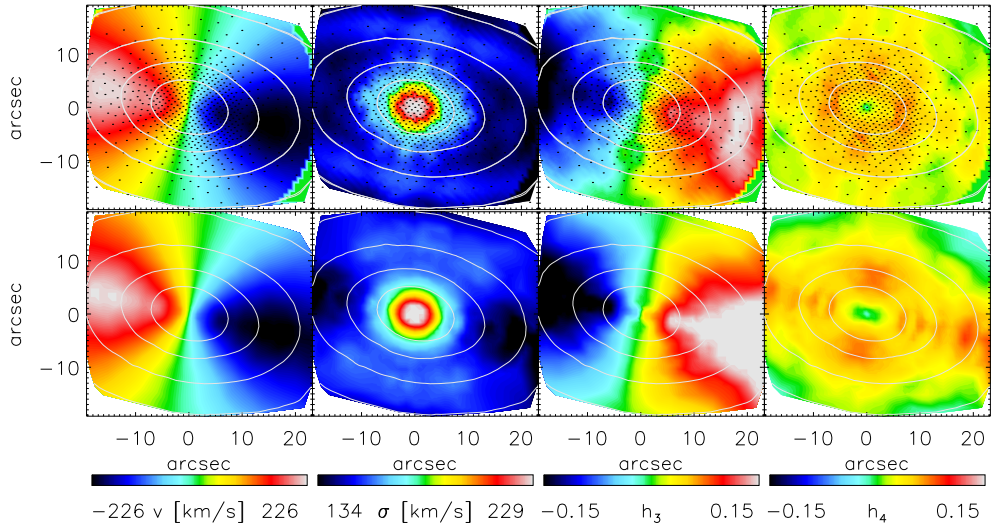, width=1.\textwidth}
 \caption{Symmetrized {\tt SAURON} data of NGC~2974 and the best
   fitting dynamical model; from right to left: stellar mean velocity
   ($v$), velocity dispersion ($\sigma$), Gauss-Hermite moments
   $h_{3}$ and $h_{4}$. First row: {\tt SAURON} data filtered by means
   of Fourier expansion with enforced bisymmetry as a visual aid for
   comparison of the data to the axisymmetric model. Original Voronoi
   2D-binned data used to constrain the model can be found in Emsellem
   et al. (2003). Centroids of the bins are shown with dots. Second
   row: prediction from the best fitting regularized orbit
   superposition dynamical model at $i = 60^{\circ}$ and $M/L = 3.9$
   (I-band). }
\end{figure}

\end{document}